# Learning Structural Node Embeddings via Diffusion Wavelets

Claire Donnat, Marinka Zitnik, David Hallac, Jure Leskovec
Stanford University
{cdonnat,marinka,hallac,jure}@stanford.edu

## ABSTRACT

Nodes residing in different parts of a graph can have similar structural roles within their local network topology. The identification of such roles provides key insight into the organization of networks and can be used for a variety of machine learning tasks. However, learning structural representations of nodes is a challenging problem, and it has typically involved manually specifying and tailoring topological features for each node. In this paper, we develop GraphWave, a method that represents each node's network neighborhood via a low-dimensional embedding by leveraging heat wavelet diffusion patterns. Instead of training on hand-selected features, GraphWave learns these embeddings in an unsupervised way. We mathematically prove that nodes with similar network neighborhoods will have similar GraphWave embeddings even though these nodes may reside in very different parts of the network. GraphWave runtime scales linearly with the number of edges and experiments in a variety of different settings demonstrate GraphWave's real-world potential for capturing structural roles in networks. All in all, GraphWave outperforms existing state-of-the-art baselines in every experiment, by as much as 137%.

## CCS CONCEPTS

• **Networks** → **Topology analysis and generation**; • **Computing methodologies** → **Kernel methods**; **Learning latent representations**; **Spectral methods**; *Cluster analysis*; *Motif discovery*; • **Information systems** → *Clustering*; Nearest-neighbor search;

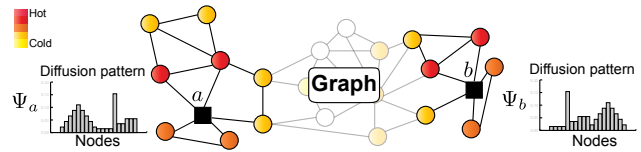

**Figure 1: Nodes $a$ and $b$ have similar structural roles even though they are distant in the graph. While the raw spectral graph wavelets of $a$ and $b$ might be very different, we treat them as probability distributions and prove that the distributions of wavelet coefficients of structurally similar nodes are indeed similar.**

## 1 INTRODUCTION

Structural role discovery in graphs focuses on identifying nodes which have topologically similar network neighborhoods while residing in potentially distant areas of the network (Figure 1). Intuitively, nodes with similar structural roles perform similar functions in the network, such as managers in the social network of a company or enzymes in the molecular network of a cell. This alternative definition of node similarity is very different than more traditional notions [9, 12–14, 20, 24, 26, 35], which assume some measure of "smoothness" over the graph and thus consider nodes residing in close network proximity to be similar. Such structural role information about the nodes can be used for a variety of tasks, including as input to machine learning problems, or even to identify key nodes in a system (principal "influencers" in a social network, critical hubs in contagion graphs, etc.).

When structural roles of nodes are defined over a discrete space, they correspond to different topologies of local network neighborhoods (*e.g.*, node on a chain, center of a star, a bridge between two clusters). However, such discrete roles must be pre-defined, requiring domain expertise and manual inspection of the graph structure. A more powerful and robust method for identifying structural similarity involves learning a continuous vector-valued *structural embedding* $\chi_a$ of each node $a$ in an unsupervised way. This motivates a natural definition of structural similarity in terms of closeness of topological embeddings: For any $\epsilon > 0$, nodes $a$ and $b$ are defined to be $\epsilon$-structurally similar with respect to a given distance if $dist(\chi_a, \chi_b) \leq \epsilon$. Thus, a robust approach must introduce both an appropriate embedding and an adequate distance metric.

While several methods have been proposed for learning structural embeddings of nodes in graphs, existing approaches are extremely sensitive to small perturbations in the topology and lack mathematical understanding of the properties of the learned embeddings. Furthermore, they often require manually hand-labeling topological features [16], rely on non-scalable heuristics [27], and/or return a single similarity score instead of a multidimensional structural embedding [18, 19].

**Present work.** Here we address the problem of structure learning on graphs by developing GraphWave. Building upon techniques from graph signal processing [5, 15, 30], our approach learns a multidimensional structural embedding for each node based on the diffusion of a spectral graph wavelet centered at the node. Intuitively, each node propagates a unit of energy over the graph and characterizes its neighboring topology based on the response of the network to this probe. We formally prove that the coefficients of this wavelet directly relate to graph topological properties. Hence, these coefficients contain all the necessary information to recover structurally similar nodes, without requiring the explicit hand-labeling of features. However, the wavelets are, by design, localized on the graph. Therefore to compare wavelets for nodes that are far away from each other, typical graph signal processing methods (using metrics like correlation between wavelets or $\ell_2$ distance) cannot be used without specifying an exact one-to-one mapping between nodes for every pairwise comparison, a computationally intractable task. For this reason, these wavelets have never before been used for learning structural embeddings.

To overcome this challenge, we propose a novel way of treating the wavelets as probability distributions over the graph. This way, the structural information is contained in *how* the diffusion spreads over the network rather than *where* it spreads. In order to provide vector-valued embeddings, we embed these wavelet distributions using the empirical characteristic function [23]. The advantage of empirical characteristic functions is that they capture all the moments (including higher-order moments) of a given distribution. This allows GraphWave to be robust to small perturbations in the local edge structure, as we prove mathematically. The computational complexity of GraphWave is linear in the number of edges, thus allowing it to scale to large (sparse) networks. Finally, we compare GraphWave to several state-of-the-art baselines on both real and synthetic datasets, obtaining improvements of up to 137% and demonstrating how our approach is a useful tool for structural embeddings in graphs.

**Summary of contributions.** The main contributions of our paper are as follows:
- We propose a novel use of spectral graph wavelets by treating them as probability distributions and characterizing the distributions using empirical characteristic functions.
- We leverage these insights to develop a scalable method (GraphWave) for learning node embeddings based on structural similarity in graphs, outperforming existing state-of-the-art baselines.
- We prove mathematically that GraphWave accurately recovers structurally similar and structurally equivalent nodes.

**Further related work.** Prior work on discovering nodes with similar structural roles has typically relied on explicit featurization of nodes. These methods generate an exhaustive listing of each node's local topological properties (*e.g.*, node degree, number of triangles it participates in, number of $k$-cliques, its PageRank score) before computing node similarities based on such heuristic representations. A notable example of such approaches is *RolX* [11, 16], a matrix-factorization based method which aims to recover a soft-clustering of nodes into a predetermined number of $K$ distinct roles using recursive feature extraction [17]. Similarly, *struc2vec* [27] uses a heuristic to construct a multilayered graph based on topological metrics and simulates random walks on the graph to capture structural information. In contrast, our approach does not rely on heuristics (we mathematically prove its efficacy) and does not require explicit manual feature engineering or hand-tuning of parameters.

Recent neural representation learning methods (structure2vec [6], neural fingerprints [8], graph convolutional networks (GCNs) [13,

20], message passing networks [10], etc.) are a related line of research. However, these graph embedding methods do not apply in our setting, since they solve a (supervised) graph classification task and/or embed entire graphs while we embed individual nodes.

Another line of related work are graph diffusion kernels [5] which have been utilized for various graph modeling purposes [3, 22, 29, 34]. However, to the best of our knowledge, our paper is the first to apply graph diffusion kernels for determining structural roles in graphs. Kernels have been shown to efficiently capture geometrical properties and have been successfully used for shape detection in the image processing community [1, 25, 33]. However, in contrast to shape-matching problems, GraphWave considers these kernels as probability distributions over real-world graphs. This is because the graphs that we consider are highly irregular (as opposed to the Euclidean and manifold graphs). Therefore, traditional wavelet methods, which typically analyze node diffusions across specific nodes that occur in regular and predictable patterns, do not apply. Instead, GraphWave characterizes the *shape* of the diffusion, rather than the specific nodes where the diffusion occurs. This key insight allows us to uncover structural embeddings and to discover structurally similar nodes.

## 2 LEARNING STRUCTURAL EMBEDDINGS

Given an undirected graph $G = (\mathcal{V}, \mathcal{E})$ with $N$ nodes $\mathcal{V} = \{a_1, \ldots, a_N\}$, edges $\mathcal{E}$, an adjacency matrix $A$ (binary or weighted), and a degree matrix $D_{ii} = \sum_j A_{ij}$, we consider the problem of learning, for every node $a_i$, a *structural embedding* representing $a_i$'s position in a continuous multidimensional space of structural roles.

We frame this as an unsupervised learning problem based on spectral graph wavelets [15] and develop an approach called GraphWave that provides mathematical guarantees on the optimality of learned structural embeddings.

### 2.1 Spectral graph wavelets

In this section, we provide background on the spectral graph wavelet-based model [15, 30] that we will use in the rest of the paper.

Let $U$ be the eigenvector decomposition of the unnormalized graph Laplacian $L = D - A = U\Lambda U^T$ and let $\lambda_1 < \lambda_2 \leq \cdots \leq \lambda_N$ ($\Lambda = \text{Diag}(\lambda_1, \ldots, \lambda_N)$) denote the eigenvalues of $L$.

Let $g_s$ be a filter kernel with scaling parameter $s$. In this paper, we use the heat kernel $g_s(\lambda) = e^{-\lambda s}$, but our results apply to any scaling wavelet [31]. For now, we assume that $s$ is given; we develop a method for selecting an appropriate value of $s$ in Section 4.

Graph signal processing [15, 30] defines the spectral graph wavelet associated with $g_s$ as the signal resulting from the modulation in the spectral domain of a Dirac signal centered around node $a$. The spectral graph wavelet $\Psi_a$ is given by an $N$-dimensional vector:

$$\Psi_a = U \, \text{Diag}(g_s(\lambda_1), \ldots, g_s(\lambda_N)) U^T \delta_a, \quad (1)$$

where $\delta_a = \mathbb{1}(a)$ is the one-hot vector for node $a$. For notational simplicity, we drop the explicit dependency of spectral graph wavelet $\Psi_a$ on $s$. The $m$-th wavelet coefficient of this column vector is thus given by $\Psi_{ma} = \sum_{l=1}^{N} g_s(\lambda_l) U_{ml} U_{al}$.

In spectral graph wavelets, the kernel $g_s$ modulates the eigenspectrum such that the resulting signal is typically *localized on the graph and in the spectral domain* [30]. Spectral graph wavelets are based on an analogy between temporal frequencies of a signal and the Laplacian's eigenvalues. Eigenvectors associated with smaller eigenvalues carry slow varying signal, encouraging nodes that are neighbors to share similar values. In contrast, eigenvectors associated with larger eigenvalues carry faster-varying signal across edges. The low-pass filter kernel $g_s$ can thus be seen as a modulation operator that discounts higher eigenvalues and enforces smoothness in the signal variation on the graph.

## 2.2 GraphWave algorithm

We first describe GraphWave (Algorithm 1); then, we analyze it in the next section. For every node $a$, GraphWave returns a $2d$-dimensional vector $\chi_a$ representing its *structural embedding*, where nodes with structurally similar local network neighborhoods will have similar embeddings.

We first apply spectral graph wavelets to obtain a diffusion pattern for every node (Line 3), which we gather in a matrix $\Psi$. Here, $\Psi$ is a $N \times N$ matrix, where $a$-th column vector is the spectral graph wavelet for a heat kernel centered at node $a$. In contrast to prior work that studies wavelet coefficients as a function of the scaling parameter $s$, we study them as a function of the network (*i.e.*, how the coefficients vary across the local network neighborhood around the node $a$). In particular, coefficients in each wavelet are identified with the nodes and $\Psi_{ma}$ represents the amount of energy that node $a$ has received from node $m$. As we will later show nodes $a$ and $b$ with similar network neighborhoods have similar spectral wavelet coefficients $\Psi$ (assuming that we know how to solve the "isomorphism" problem and find the explicit one-to-one mapping of the nodes from $a$'s neighborhood to the nodes of the $b$'s neighborhood). To resolve the node mapping problem GraphWave treats the wavelet coefficients as a probability distribution and characterizes the distribution via empirical characteristic functions. This is a key insight that makes it possible for GraphWave to learn nodes' structural embeddings via spectral graph wavelets.

More precisely, we embed spectral graph wavelet coefficient distributions into $2d$-dimensional space (Line 4-7) by calculating the characteristic function for each node's coefficients $\Psi_a$ and sample it at $d$ evenly spaced points. The characteristic function of a probability distribution $X$ is defined as: $\phi_X(t) = \mathbb{E}[e^{itX}], t \in \mathbb{R}$. The function $\phi_X(t)$ fully characterizes the distribution of $X$ because it captures information about all the moments of probability distribution $X$ [23]. For a given node $a$ and scale $s$, the empirical characteristic function of $\Psi_a$ is defined as:

$$\phi_a(t) = \frac{1}{N} \sum_{m=1}^{N} e^{it\Psi_{ma}} \quad (2)$$

Finally, structural embedding $\chi_a$ of node $a$ is obtained by sampling the 2-dimensional parametric function (Eq. (2)) at $d$ evenly spaced points $t_1, \ldots t_d$ and concatenating the values:

$$\chi_a = \left[ \text{Re}(\phi_a(t_i)), \text{Im}(\phi_a(t_i)) \right]_{t_1, \cdots t_d} \quad (3)$$

Note that we sample the empirical characteristic function $\phi_a(t)$ at $d$ points, which creates a structural embedding of size $2d$, so the dimensionality of the embedding is independent of the graph size.

**Distance between structural embeddings.** The output of Graph-Wave is a structural embedding $\chi_a$ for each node $a$ in the graph.

---

**Algorithm 1** Learning structural embeddings in GraphWave.

1: **Input:** Graph $\mathcal{G} = (\mathcal{V}, \mathcal{E})$, scale $s$, evenly spaced sampling points $\{t_1, t_2, \ldots, t_d\}$
2: **Output:** Structural embedding $\chi_a \in \mathbb{R}^{2d}$ for every node $a \in \mathcal{V}$
3: Compute $\Psi = U g_s(\Lambda) U^T$ (Eq. (1))
4: **for** $t \in \{t_1, t_2, \ldots, t_d\}$ **do**
5: $\quad$ Compute $\phi(t) = $ column-wise mean$(e^{it\Psi}) \in \mathbb{R}^N$
6: $\quad$ **for** $a \in \mathcal{V}$ **do**
7: $\quad\quad$ Append $\text{Re}(\phi_a(t))$ and $\text{Im}(\phi_a(t))$ to $\chi_a$

---

We can explore distances between these embeddings through the use of the $\ell_2$ distance on $\chi_a$. The structural distance between nodes $a$ and $b$ is then defined as: $dist(a, b) = \|\chi_a - \chi_b\|_2$. By definition of the characteristic function, this amounts to comparing moments of different orders defined on wavelet coefficient distributions.

**Scaling parameter.** The scaling parameter $s$ determines the radius of network neighborhood around each node $a$ ([15, 34]). A small value of $s$ determines node embeddings based on similarity of nodes' immediate neighborhoods. In contrast, a larger value of $s$ allows the diffusion process to spread farther in the network, resulting in embeddings based on neighborhoods with greater radii.

GraphWave can also integrate information across different radii of neighborhoods by jointly considering many different values of $s$. This is achieved by concatenating $J$ representations $\chi_a^{(s_j)}$, each associated with a scale $s_j$, where $s_j \in [s_{\min}, s_{\max}]$. We provide a theoretically justified method for finding an appropriate range $s_{\min}$ and $s_{\max}$ in Section 4. In this multiscale version of GraphWave, the final aggregated structural embedding for node $a$ is a vector $\chi_a \in \mathbb{R}^{2dJ}$ with the following form: $\chi_a = [\text{Re}(\phi_a^{(s_j)}(t_i)), \text{Im}(\phi_a^{(s_j)}(t_i))]_{t_i, s_j}$.

**Computational complexity.** We use Chebyshev polynomials [32] to compute Line 3 in Algorithm 1. As in [7], each power of the Laplacian has a computational cost of $O(|\mathcal{E}|)$, yielding an overall complexity of $O(K|\mathcal{E}|)$, where $K$ denotes the order Chebyshev polynomial approximation. The overall complexity of GraphWave is linear in the number of edges, which allows GraphWave to scale to large sparse networks.

## 3 ANALYSIS OF GRAPHWAVE

In this section, we provide theoretical motivation for our spectral graph wavelet-based model. First we analytically show that spectral graph wavelet coefficients characterize the topological structure of local network neighborhoods (Section 3.1). Then we show that structurally equivalent/similar nodes have near-identical/similar embeddings (Sections 3.2 and 3.3), thereby providing a mathematical guarantee on the optimality of GraphWave.

### 3.1 Network structure via diffusion wavelets

We start by establishing the relationship between the spectral graph wavelet of a given node $a$ and the topological properties of local network neighborhood centered at $a$. In particular, we prove that a wavelet coefficient $\Psi_{ma}$ provides a measure of network connectivity between nodes $a$ and $m$.

We use the fact that the spectrum of the graph Laplacian is discrete and contained in the compact set $[0, \lambda_N]$. It then follows from the Stone-Weierstrass theorem that the restriction of kernel $g_s$ to the interval $[0, \lambda_N]$ can be approximated by a polynomial. This polynomial approximation, denoted as $P$, is tight and its error can be uniformly bounded. Formally, this means:

$$\forall \epsilon > 0, \quad \exists P : P(\lambda) = \sum_{k=0}^{K} \alpha_k \lambda^k \quad (4)$$
$$\text{such that} \quad |g_s(\lambda) - P(\lambda)| \leq \epsilon \quad \forall \lambda \in [0, \lambda_{\max}],$$

where $K$ is the order of polynomial approximation, $\alpha_k$ are coefficients of the polynomial, and $r(\lambda) = g_s(\lambda) - P(\lambda)$ is the residual. We can now express the spectral graph wavelet for node $a$ in terms of the polynomial approximation as:

$$\Psi_a = \left(\sum_{k=0}^{K} \alpha_k L^k\right) \delta_a + U r(\Lambda) U^T \delta_a. \quad (5)$$

We note that $\Psi_a$ is a function of $L^k = (D - A)^k$ and thus can be interpreted using graph theory. In particular, it contains terms of the form $D^k$ (capturing the degree), $A^k$ (capturing the number of $k$-length paths that node $a$ participates in), and terms containing both $A$ and $D$, which denote paths of length up to $k$ going from node $a$ to every other node $m$.

Using the Cauchy-Schwartz's inequality and the facts that $U$ is unitary and $r(\lambda)$ is uniformly bounded (Eq. (4)), we can bound the second term on the right-hand side of Eq. (5) by:

$$|\delta_m^T U r(\Lambda) U^T \delta_a|^2 \leq \left(\sum_{j=1}^{N} |r(\lambda_j)|^2 U_{aj}^2\right)\left(\sum_{j=1}^{N} U_{mj}^2\right) \leq \epsilon^2. \quad (6)$$

As a consequence, each wavelet $\Psi_a$ can be approximated by a $K$-th order polynomial that captures information about the $K$-hop neighborhood of node $a$. The analysis of Eq. (5), where we show that the second term is limited by $\epsilon$, indicates that spectral graph wavelets are predominately governed by topological features (specifically, degrees, cycles and paths) according to the specified heat kernel. The wavelets thus contain the information necessary to generate structural embeddings of nodes.

### 3.2 Embeddings of structurally equivalent nodes

Let us consider nodes $a$ and $b$ whose $K$-hop neighborhoods are identical (where $K$ is an integer less than the diameter of the graph), meaning that nodes $a$ and $b$ are structurally equivalent. We now show that $a$ and $b$ have $\epsilon$-structurally similar embeddings in GraphWave.

First, we use the Taylor expansion to obtain an explicit $K$-th order polynomial approximation of $g_s$ as $P(\lambda, s) = \sum_{k=0}^{K} (-1)^k (s\lambda)^k / k!$. Then, for each eigenvalue $\lambda$, we use the Taylor-Lagrange equality to ensure the existence of $c_\lambda \in [0, s]$ such that:

$$|r(\lambda)| = |e^{-\lambda s} - P(\lambda, s)| = \frac{(\lambda s)^{K+1}}{(K+1)!} e^{-\lambda c_\lambda} \leq \frac{(\lambda s)^{K+1}}{(K+1)!}. \quad (7)$$

If we take any $s$ that satisfies: $s \leq ((K+1)!\epsilon))^{1/(K+1)}/\lambda_2$, then the absolute residual $|r(\lambda)|$ in Eq. (7) can be bounded by $\epsilon$ for each eigenvalue $\lambda$. Here, $\epsilon$ is a parameter that we can specify depending on how close we want the embeddings of structurally equivalent nodes to be (note that smaller values of the scale $s$ lead to smaller values of $\epsilon$ and thus tighter bounds).

Because $a$ and $b$ are structurally equivalent, there exists a one-to-one mapping $\pi$ from the $K$-hop neighborhood of $a$ (i.e., $\mathcal{N}_K(a)$) to the $K$-hop neighborhood of $b$ (i.e., $\mathcal{N}_K(b)$), such that: $\mathcal{N}_K(b) = \pi(\mathcal{N}_K(a))$. We extend the mapping $\pi$ to the whole graph $\mathcal{G}$ by randomly mapping the remaining nodes. Using Eq. (5), we write the difference between each pair of mapped coefficients $\Psi_{ma}$ and $\Psi_{\pi(m)b}$ in terms of the $K$-th order approximation of the graph Laplacian:

$$|\Psi_{ma} - \Psi_{\pi(m)b}| = \left|\delta_m U(P(\Lambda) + r(\Lambda))U^T \delta_a - \right.$$
$$\left.\delta_{\pi(m)} U(P(\Lambda) + r(\Lambda))U^T \delta_b\right| \leq \left|(UP(\Lambda)U^T)_{ma} - (UP(\Lambda)U^T)_{\pi(m)a}\right|$$
$$+ \left|(Ur(\Lambda)U^T)_{ma}\right| + \left|(Ur(\Lambda)U^T)_{\pi(m)b}\right|. \quad (8)$$

Here, we analyze the first term on the second line in Eq. (8). Since the $K$-hop neighborhoods around $a$ and $b$ are identical and by the localization properties of the $k$-th power of the Laplacian ($k$-length paths, Section 3.1), the following holds:

$$\forall m \in \mathcal{N}_K(a), \left(\sum_{k=0}^{K} \alpha_k L^k\right)_{ma} = \left(\sum_{k=0}^{K} \alpha_k L^k\right)_{\pi(m)b},$$

$$\forall m \notin \mathcal{N}_K(a), \left(\sum_{k=0}^{K} \alpha_k L^k\right)_{ma} = \left(\sum_{k=0}^{K} \alpha_k L^k\right)_{\pi(m)a} = 0,$$

meaning that this term cancels out in Eq. (8). To analyze the second and third terms on the second line of Eq. (8), we use the bound for the residual term in the spectral graph wavelet (Eq. (6)) to uniformly bound entries in matrix $Ur(\Lambda)U^T$ by $\epsilon$.

Therefore, each wavelet coefficient in $\Psi_a$ is within $2\epsilon$ of its corresponding wavelet coefficient in $\Psi_b$, i.e., $|\Psi_{ma} - \Psi_{\pi(m)b}| \leq 2\epsilon$.

As a result, because similarity in distributions translates to similarity in the resulting characteristic functions (Lévy's continuity theorem), then assuming the appropriate selection of scale, structurally equivalent nodes have $\epsilon$-structurally similar embeddings.

### 3.3 Embeddings of structurally similar nodes

We now analyze structurally similar nodes, or nodes whose $K$-hop neighborhoods are identical up to a small perturbation of the edges. We show that such nodes have similar GraphWave embeddings.

Let $\widetilde{\mathcal{N}}_K(a)$ denote a perturbed $K$-hop neighborhood of node $a$ obtained by rewiring edges in the original $K$-hop neighborhood $\mathcal{N}_K(a)$. We denote by $\tilde{L}$ the graph Laplacian associated with that perturbation. We next show that when perturbation of a node neighborhood is small, the changes in node's wavelet coefficients are small as well.

Formally, assuming a small perturbation of the graph structure (i.e., $\sup \|L^k - \tilde{L}^k\|_F \leq \epsilon$, for all $k \leq K$), we use $K$-th order Taylor expansion of kernel $g_s$ to express the wavelet coefficients in the perturbed graph as:

$$\widetilde{\Psi}_a = \sum_{k=0}^{K} \alpha_k \tilde{L}^k + \tilde{U} r(\tilde{\Lambda}) \tilde{U}^T. \quad (9)$$

We then use the Weyl's theorem [4] to relate perturbations in the graph structure to the change in the eigenvalues of the graph Laplacian. In particular, a small perturbation of the graph yields small perturbations of the eigenvalues. That is, for each $\tilde{\lambda}$, $r(\tilde{\lambda})$ is close its original value $r(\lambda)$: $r(\tilde{\lambda}) = r(\lambda) + o(\epsilon) \leq C\epsilon$, where $C$ is a constant. Taking everything together, we get:

$$|\Psi_{ma} - \widetilde{\Psi}_{ma}| \leq |\sum_{k=0}^{K} \alpha_k (L^k - \tilde{L}^k)_{ma}| + |\tilde{U}r(\tilde{\Lambda})\tilde{U}^T|_{ma}$$
$$+ |Ur(\Lambda)U^T|_{ma} = (\sum_{k=0}^{K} |\alpha_k| + 1 + C)\epsilon,$$

indicating that structurally similar nodes have similar embeddings in GraphWave.

## 4 SCALE OF HEAT DIFFUSION WAVELETS

Here, we develop a method that automatically finds an appropriate range of values for the scaling parameter $s$ in heat kernel $g_s$, which we use in the multiscale version of GraphWave (Section 2.2).

We do so by specifying an interval bounded by $s_{\min}$ and $s_{\max}$ through the analysis of variance in heat diffusion wavelets. Intuitively, small values of $s$ allow little time for the heat to propagate, yielding diffusion distributions (i.e., heat diffusion wavelet distributions) that are trivial in the sense that only a few coefficients have non-zero values and are thus unfit for comparison. For larger values of $s$, the network converges to a state in which all nodes have an identical temperature equal to $1/N$, meaning that diffusion distributions are data-independent, hence non-informative.

Here, we prove two propositions to provide new insights into the variance and convergence rate of heat diffusion wavelets. We then use these results to select $s_{\min}$ and $s_{\max}$.

PROPOSITION 1. *The variance of off-diagonal coefficients in heat diffusion wavelet $\Psi_a^{(s)}$ is proportional to:*

$$\mathrm{Var}[\{\Psi_{am}^{(s)}; m \neq a\}] \propto \Delta_a^{(0)} \Delta_a^{(2s)} - (\Delta_a^{(s)})^2,$$

*where $\Delta_a^{(s)} = |\Psi_{aa}^{(s)} - \frac{1}{N}|$ decreases monotonically with $s$.*

PROOF. Let us denote the mean of off-diagonal coefficients in wavelet $\Psi_a$ by $\tilde{\mu}_a^{(s)} = \sum_{m \neq a} \Psi_{ma}^{(s)}/(N-1)$. We use the fact that $\sum_{m \neq a} \Psi_{ma}^{(s)} = 1 - \Psi_{aa}^{(s)}$, along with the definition of the variance, to obtain:

$$\mathrm{Var}[\{\Psi_{am}^{(s)}; m \neq a\}] = \frac{1}{N-1} \sum_{m \neq a} (\Psi_{ma}^{(s)} - \tilde{\mu}_a^{(s)})^2$$
$$= \frac{1}{N-1} \sum_{m \neq a} (\Psi_{ma}^{(s)})^2 - (\tilde{\mu}_a^{(s)})^2$$
$$= \frac{N}{(N-1)^2}(\Psi_{aa}^{(2s)}\frac{N-1}{N} - (\Psi_{aa}^{(s)})^2 + \frac{2\Psi_{aa}^{(s)}}{N} - \frac{1}{N})$$
$$= \frac{N}{(N-1)^2}(\Delta_a^{(0)}\Delta_a^{(2s)} - (\Delta_a^{(s)})^2).$$

□

Proposition 1 proves that the variance is a function of $\Delta_a^{(s)}$. Therefore, to maximize the variance, we must analyze the behavior of $\Delta_a^{(s)}$. To ensure sufficient variability in the distribution of wavelet coefficients, we need to select a range $[s_{min}, s_{max}]$ that bounds $\Delta_a^{(s)}$. Our goal thus becomes establishing that $\Delta_a^{(s)}$ is large enough that the diffusion has had time to spread, while remaining sufficiently small to ensure that the diffusion is far from its converged state.

PROPOSITION 2. *The convergence of heat diffusion wavelet coefficient $\Psi_{am}^{(s)}$ is bounded by:*

$$e^{-\lambda_N \lceil s \rceil} \Delta_a^{(0)} \leq \Delta_a^{(s)} \leq e^{-\lambda_2 \lfloor s \rfloor} \Delta_a^{(0)}.$$

PROOF. For non-negative $s$, $|\Psi_{aa}^{(s+1)} - \frac{1}{N}| = |\sum_{j=2}^{N} e^{-\lambda_j(s+1)} U_{ja}^2| \leq e^{-\lambda_2}|\Psi_{aa}^{(s)} - \frac{1}{N}|$, and, symetrically, $|\Psi_{aa}^{(s+1)} - \frac{1}{N}| \geq e^{-\lambda_N}|\Psi_{aa}^{(s)} - \frac{1}{N}|$. We conclude that $e^{-\lambda_N} \leq |\Psi_{aa}^{(s+1)} - \frac{1}{N}|/|\Psi_{aa}^{(s)} - \frac{1}{N}| \leq e^{-\lambda_2}$. Given any $s \in \mathbb{N}$, we use the induction principle to get $e^{-\lambda_N s}|\Psi_{aa}^{(0)} - \frac{1}{N}| \leq |\Psi_{aa}^{(s)} - \frac{1}{N}| \leq e^{-\lambda_2 s}|\Psi_{aa}^{(0)} - \frac{1}{N}|$, which yields the desired bound, $e^{-\lambda_N s}\Delta_a^{(0)} \leq \Delta_a^{(s)} \leq e^{-\lambda_2 s}\Delta_a^{(0)}$. Since $\Delta_a^{(s)}$ is a smooth increasing function of $s$, we take the floor/ceiling of any non-integer $s \geq 0$ and this proposition must hold. □

**Selection of $s_{\max}$.** We select $s_{\max}$ such that wavelet coefficients are localized in the network. To do so, we use Proposition 2 and bound $\Delta_a^{(s)}$ by the graph Laplacian's eigenvalues. When the bulk of the eigenvalues leans towards $\lambda_N$, $\Delta_a^{(s)}$ is closer to $e^{-\lambda_N}$ (i.e., lower bound in Proposition 2). When the bulk of the eigenvalues is closer to $\lambda_2$, $\Delta_a^{(s)}$ will lean towards $e^{-\lambda_2}$ (i.e., upper bound in Proposition 2). In each case, the diffusion is localized if $\Delta_a^{(s)}$ is above a given threshold $\eta < 1$. Indeed, this ensures that $\Delta_a^{(s)}$ has shrunk to at most $\eta * 100$ % of its initial value at $s = 0$, and yields a bound of the form

$\Delta_a^{(s)}/\Delta_a^{(0)} \geq \eta$. The bound implies that: $e^{-\lambda s} \geq \eta$, or $s \leq -\log(\eta)/\lambda$. To find a middle ground between the two convergence scenarios, we take $\lambda$ to be the geometric mean of $\lambda_2$ and $\lambda_N$. As opposed to the arithmetic mean, the geometric mean maintains an equal weighting across the range $[\lambda_2, \lambda_N]$, and a change of $\epsilon\%$ in $\lambda_2$ has the same effect as a change in $\epsilon\%$ of $\lambda_N$. We thus select $s_{\max}$ as $-\log(\eta)/\sqrt{\lambda_2 \lambda_N}$.

**Selection of $s_{\min}$.** We select $s_{\min}$ to ensure the adequate diffusion resolution. In particular, we select a minimum value $s_{\min}$ such that each wavelet has sufficient time to spread. That is, $\Delta_a^{(s)}/\Delta_a^{(0)} \leq \gamma$. As in the case of $s_{\max}$ above, we obtain a bound of $s \geq -\log(\gamma)/\lambda$. Hence, we set $s_{\min}$ to $-\log(\gamma)/\sqrt{\lambda_2 \lambda_N}$.

To cover an appropriate range of scales, we suggest setting $\eta = 0.85$ and $\gamma = 0.95$.

## 5 EXPERIMENTS ON SYNTHETIC GRAPHS

GraphWave's embeddings are independent of any downstream task, so we evaluate them in a variety of different synthetic settings to demonstrate their potential for capturing structural roles in networks.

**Baseline methods.** We evaluate the performance of GraphWave[1] against two state-of-the-art baselines for learning structural embeddings: *struc2vec* [27], a method which discovers structural embeddings at different scales through a sequence of walks on a multilayered graph, and *RolX* [16], a method based on non-negative matrix factorization of a node-feature matrix (number of neighbors, triangles, etc.) that describes each node based on this given set of latent features. While in [16], the authors develop a method for automatically selecting the number of roles in *RolX*, we use *RolX* as an oracle estimator, providing it with the correct number of classes. We note that GraphWave and *struc2vec* learn embeddings on a continuous spectrum instead of into discrete classes (and thus they do not require this parameter).

---
[1]All code can be downloaded at: http://snap.stanford.edu/graphwave.

We also compare GraphWave with two recent unsupervised node representation learning methods, *node2vec* [12] and *DeepWalk* [26], to emphasize the difference between such methods and our structural similarity-based approach. For all baselines, we use the default parameter values in the available solvers, and for GraphWave, we use the multiscale version (Section 4), set $d = 50$ and use evenly spaced sampling points $t_i$ in range $[0, 100]$. We again note that graph embedding methods (structure2vec [6], neural fingerprints [8], GCNs [13, 20], etc.) do not apply in these settings, since they embed entire graphs while we embed individual nodes.

## 5.1 Barbell graph

We first consider a barbell graph consisting of two dense cliques connected by a long chain (Figure 2A). We plot a 2D PCA representation the learned embeddings of the three structural-based methods, GraphWave, *RolX*, and *struc2vec*, in Figures 2B-D.

GraphWave correctly learns identical representations for structurally equivalent nodes, providing empirical evidence for our theoretical result in Section 3.2. This can be seen by structurally equivalent nodes in Figure 2A (nodes of the same color) having identical projections and overlap in the PCA plot (Figure 2D). In contrast, both *RolX* and *struc2vec* fail to recover the exact structural equivalences (as shown by the non-overlapping nodes of the same color). For *struc2vec*, the projections are not consistent with the expected ordering: the yellow nodes are farther to the green ones than the red. In contrast, *RolX* yields highly-similar node embeddings whose projections cluster into three high-level groups, meaning that *RolX* can identify three out of eight structural classes in the barbell graph.

We also note that all three methods correctly group the clique nodes (purple) together. However, only GraphWave correctly differentiates between nodes connecting the two dense cliques in the barbell graph, providing empirical evidence for our theoretical result in Section 3.3. GraphWave represents those nodes in a gradient-like pattern that captures the spectrum of structural roles of those nodes (Figure 2D).

## 5.2 Graphs with planted structural equivalences

We next systematically evaluate the methods on synthetic graphs. We develop a procedure that can generate a graph with planted structural equivalences and also ground-truth node labels indicating the structural role of each node. Our goal is to use these ground-truth structural roles to evaluate our method's performance.

**Generating the graphs.** The graphs are given by basic shapes of one of different types ("house", "fan", "star") that are regularly placed along a cycle (Table 1 and Figure 3A) of length 30. In the "varied" setup, we mix the three basic shapes by placing 8 instances of each type randomly along a cycle (of length 40), thus generating synthetic graphs with richer and more complex structural role patterns. Additional "noisy" graphs are generated by adding edges uniformly at random on these graphs. In our experiments, we set this number to be 10% of the edges in the original structure. This setup is designed to assess the robustness of the methods to data perturbations ("house perturbed", "varied perturbed").

**Experimental setup.** For each trial, we generate one instance of each of these four types of structure, on each of which we run the different methods to learn each embedding. Each experiment was repeated 25 times, and the performance of each algorithm was assessed by considering two settings:

- **Unsupervised setting:** We assess the ability of each method to embed close together nodes with the same ground-truth structural role. We use agglomerative clustering (with single linkage) to cluster embeddings learned by each method and evaluate the clustering quality via: (1) *homogeneity* [28], conditional entropy of ground-truth structural roles given the predicted clustering; (2) *completeness* [28], a measure of how many nodes with the same ground-truth structural role are assigned to the same cluster, and (3) *silhouette score*, a measure of intra-cluster distance vs. inter-cluster distance.
- **Supervised setting:** We assess the performance of learned embeddings for node classification. Using 10-fold cross validation, we predict the structural role (label) of each node in the test set based on its 4-nearest neighbors in the training set as determined by the embedding space (Section 2.2) The reported score is then the average accuracy and $F_1$-score over 25 trials.

**Results.** For both the unsupervised and supervised settings, GraphWave outperforms the other four methods (Table 1). *Node2vec* and *DeepWalk* perform very poorly in all these experiments, yielding significantly lower scores than the structural equivalence-based methods. This is due to the fact that the crux of these methods is to learn node embeddings that reflect their distances in the graph. Of the remaining methods, GraphWave consistently outperforms *struc2vec*, obtaining a higher score in every metric under every setting. On average, compared to *struc2vec*, GraphWave has a 63% higher homogeneity score, 61% higher completeness score, 137% higher silhouette score, 46% higher prediction accuracy, and 51% higher $F_1$ score. Both GraphWave and *RolX* achieved perfect performance in the noise-free "house" setting. However, while *RolX* exhibits slightly higher homogeneity and completeness in two of the experiments ("house perturbed" and "varied"), GraphWave has a higher silhouette score, prediction accuracy, and $F_1$ score than *RolX* in all four experiments. Furthermore, in the most challenging setting ("varied perturbed"), GraphWave outperforms *RolX* in every metric, including by 9% in homogeneity, 8% in completeness, and 23% in silhouette score. This provides empirical evidence for our analytical result that GraphWave is robust to noise in the edge structure. The silhouette scores also show that the clusters recovered by GraphWave tend to be denser and better separated than for the other methods.

**Visualizing the embeddings.** We also highlight the visualization power of the embeddings that we recover: their associated parametric curves provide a way of visualizing differences between nodes. As an example, using cycle graph with attached "house" (Figure 3A), we plot a 2D PCA projections of GraphWave's embeddings in Figure 3B, confirming that GraphWave accurately distinguishes between nodes with distinct structural roles. We also visualize the resulting characteristic functions (Eq. (2)) in Figure 3C. In general, using characteristic function theory [23], their interpretation is:
- Nodes located in the periphery of the graph struggle to diffuse the signal over the graph, and thus span wavelets that are characterized by a smaller number of non-zero coefficients. Characteristic functions of such nodes thus span a small loop-like 2D curve.
- Nodes located in the core (dense region) of the graph tend to diffuse the signal farther away and reach farther nodes for the same value of $t$. Characteristic functions of such nodes thus have a farther projection on the $x$ and $y$ axis.

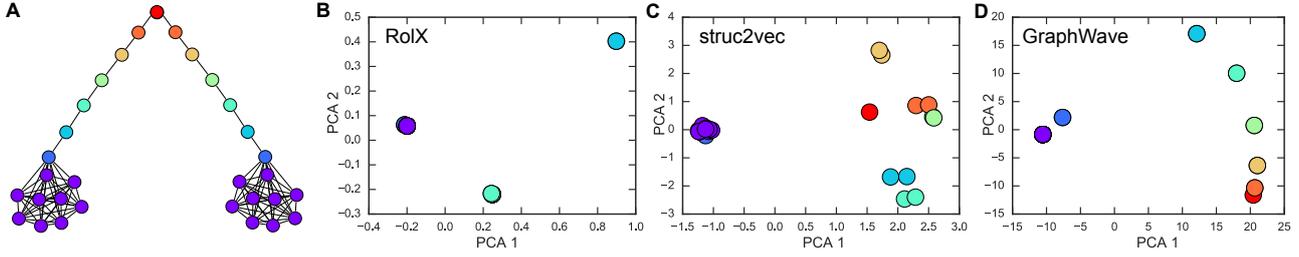

Figure 2: Barbell graph. The graph has 8 distinct classes of structurally equivalent nodes as indicated by color (A). 2D PCA projection of structural embeddings as learned by *RolX* (B), *struc2vec* (C) and GraphWave (D). Projections in (B)-(D) contain the same number of points as there are nodes in the graph (A). Identical embeddings have identical projections, resulting in overlapping points in (B)-(D).

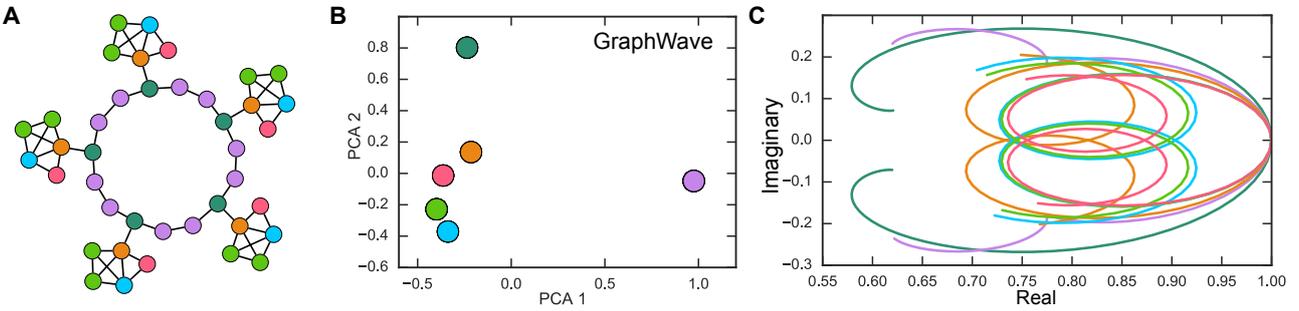

Figure 3: A cycle graph with attached "house" shapes (A). 2D PCA projection of GraphWave's embeddings. Embeddings of structurally equivalent nodes overlap, and GraphWave perfectly recovers the 6 different structural roles (B). Characteristic function for the distribution of the wavelet coefficients (C). Color of a node/curve indicates structural role. (Best seen in color.)

Table 1: Structural role discovery results for different synthetic graphs. (Best seen in color.) Results averaged over 25 synthetically generated graphs. Dashed lines denote perturbed versions of the basic shapes (obtained by randomly adding and removing edges), node colors indicate structural roles. Two best methods are shown in bold.

| Shapes placed along a cycle graph | | Method | Homogeneity | Completeness | Silhouette | Accuracy | $F_1$-score |
|---|---|---|---|---|---|---|---|
| House | 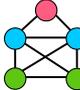 | DeepWalk | 0.002 | 0.002 | 0.29 | 0.132 | 0.107 |
| | | node2vec | 0.005 | 0.005 | 0.330 | 0.077 | 0.064 |
| | | RolX | **1.000** | **1.000** | **1.000** | **1.000** | **1.000** |
| | | struc2vec | 0.995 | 0.995 | 0.451 | 0.992 | 0.991 |
| | | GraphWave | **1.000** | **1.000** | **1.000** | **1.000** | **1.000** |
| House perturbed | 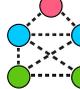 | DeepWalk | 0.059 | 0.063 | 0.247 | 0.097 | 0.081 |
| | | node2vec | 0.030 | 0.032 | 0.276 | 0.058 | 0.046 |
| | | RolX | **0.570** | **0.588** | 0.346 | **0.823** | **0.818** |
| | | struc2vec | 0.206 | 0.235 | 0.180 | 0.461 | 0.441 |
| | | GraphWave | **0.547** | **0.566** | **0.374** | **0.866** | **0.866** |
| Varied | 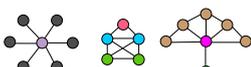 | DeepWalk | 0.262 | 0.233 | 0.354 | 0.463 | 0.428 |
| | | node2vec | 0.244 | 0.216 | 0.400 | 0.460 | 0.429 |
| | | RolX | **0.841** | **0.862** | 0.736 | 0.836 | 0.836 |
| | | struc2vec | 0.629 | 0.578 | 0.240 | 0.571 | 0.555 |
| | | GraphWave | **0.828** | **0.852** | **0.816** | **0.839** | **0.837** |
| Varied perturbed | 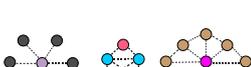 | DeepWalk | 0.298 | 0.267 | 0.327 | 0.414 | 0.387 |
| | | node2vec | 0.303 | 0.265 | 0.360 | 0.411 | 0.386 |
| | | RolX | **0.638** | **0.627** | 0.418 | 0.718 | 0.714 |
| | | struc2vec | 0.457 | 0.433 | 0.289 | 0.490 | 0.470 |
| | | GraphWave | **0.697** | **0.680** | **0.516** | **0.731** | **0.724** |

Table 2: Generalization of embeddings across graphs.

| Method | Accuracy | $F_1$-score |
|---|---|---|
| DeepWalk | 0.531 | 0.506 |
| node2vec | 0.417 | 0.369 |
| RolX | 0.868 | 0.863 |
| struc2vec | 0.767 | 0.758 |
| GraphWave | **0.936** | **0.936** |

### 5.3 Generalization of embeddings across graphs

Next we analyze a separate use case, showing how structural embeddings can identify structural similarities of nodes across different graphs. Our goal is to evaluate embeddings for their ability to transfer structural roles from nodes in one graph to nodes in another graph. That is, we test if nodes with the same structural role get embedded close together, even if they come from different graphs.

**Generating the graphs.** We generate 200 graphs with ground-truth labels for structurally equivalent nodes using the following procedure. First, we determine a skeleton of a graph (either a cycle or linear path, each with probability 0.5), then we select the size of that skeleton (i.e., cycle size or path length, uniformly at random), and finally, we attach a random number of small shapes to the skeleton (5-node house or 5-node chain, each with probability 0.5). To test the methods in noisier settings, we also randomly add 10 random edges between nodes in the basis. We use the different methods to compute embeddings for every node in each graph and then use the learned embeddings to predict each node's ground-truth structural role, for example, the corner of a house. To evaluate method's ability to generalize across graphs, we use 10-fold cross validation and predict the label of each test node in a given graph based on its 4-nearest neighbors (as defined in Section 2.2) in all other graphs. We then measure the average accuracy and $F_1$-score.

**Results.** GraphWave outperforms all alternative methods on both classification metrics (Table 2). We see that recent unsupervised node representation learning methods perform poorly, and that the second best performing method is *RolX*. However, GraphWave outperforms both *RolX* (by 8% in $F_1$-score and 8% in accuracy), and *struc2vec* (by 23% in $F_1$-score and 22% in accuracy). This highlights GraphWave's ability to learn structural signatures which are meaningful across different graphs, with high predictive power.

### 5.4 Scalability and sensitivity to noise

We next analyze scalability of GraphWave and its sensitivity to noise in input graphs.

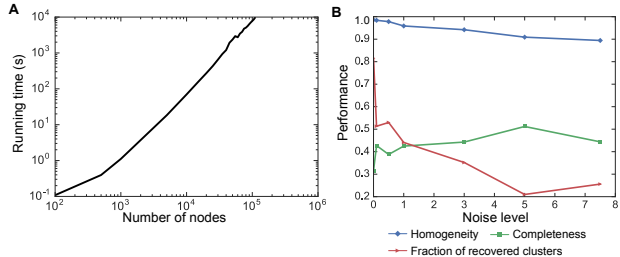

Figure 4: (A) Running time of GraphWave (in seconds) as a function of the number of nodes in an input graph. (B) Performance of GraphWave as a function of the noise level. Noise level is given by the percentage of initial number of edges that are randomly rewired. Results are averaged across ten runs.

**Scalability.** To evaluate how GraphWave scales to large graphs, we generate synthetic graphs of increasing size. Each graph has a skeleton given by a cycle graph to which we attach a number of network motifs, as described in Section 5.3. We take each graph and run GraphWave to learn structural node embeddings. Figure 4A shows GraphWave's running time as a function of the number of nodes in the graph. As the computation of scaling wavelet coefficients takes linear time in the number of edges, GraphWave lends itself to a fast algorithm that can be efficiently implemented using sparse matrix representation. A potential bottleneck of GraphWave is the conversion of the distribution induced by these wavelets into an Euclidean space via an empirical characteristic function (Section 2.2). To efficiently evaluate the characteristic function, GraphWave leverages the sparsity of the wavelet's coefficients. Note that wavelet coefficients are sparse because of GraphWave's approach to the selection of scales, which does not allow the signal to propagate too far away over the network (Section 4). This approach reduces the computation of embeddings to a set of sparse matrix multiplications and applications of element-wise functions on the non-zero elements of sparse matrices. In contrast, *struc2vec* does not scale to large graphs, as it takes as long as 260 seconds on a network of only 100 nodes, whereas GraphWave can scale to graphs of comparable sizes.

**Sensitivity to noise.** We next evaluate GraphWave's sensitivity to noise in an input graph. The noise is injected into the graph in the same manner as in the "varied perturbed" setting in Table 1. For each graph, we first learn structural node embeddings using GraphWave, then we cluster the embeddings using affinity propagation clustering algorithm, and finally, we evaluate the quality of clusters against ground-truth structural roles. Note that affinity propagation does not need the number of clusters as input but produces the clusters automatically. This is important in this experiment as the meaning of each cluster can be hindered because of high levels of noise. In addition to homogeneity and completeness metrics, we also report the number of detected clusters (as a fraction of the maximum number of possible clusters, that is, the number of nodes $N$) as a measure of the richness of roles recovered by GraphWave. Figure 4B shows performance of GraphWave as we vary the level of noise added to the input graph. Results show that GraphWave's performance degrades gracefully, even in the presence of strong noise.

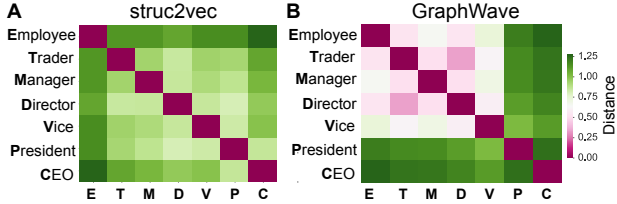

Figure 5: Heat maps indicate average distance between roles in the Enron email graph, as determined by *struc2vec* (A) and GraphWave (B).

## 6 EXPERIMENTS ON REAL-WORLD GRAPHS

We proceed by applying GraphWave to three real-world networks.

### 6.1 Mirrored Karate network

We first consider a mirrored Karate network, using the same experimental setup as in [27]. The mirrored Karate network is created by taking two copies of Zachary's karate network and adding a given number of random edges, each edge connecting mirrored nodes from different copies representing the same individual in the karate club (*i.e.*, mirrored edges). The goal of the experiment is to identify mirrored nodes by capturing structural similarity while varying the number of mirrored edges in the network.

We measure the percentage of nodes whose nearest neighbor corresponds to its true structural equivalent, comparing GraphWave with the various baselines. As the number of mirrored edges varies from 1 to 25, the results are remarkably consistent. GraphWave achieves an average accuracy of 83.2% (with a minimum score of 75.2% and a maximum score of 86.2%), *RolX* achieves an average of 82.2% (min/max of 79.4% and 84.5%, respectively), *struc2vec* scores 52.5% (43.3% and 59.4%). Alternatively, both *node2vec* and *DeepWalk* never score higher than 8.5%, since they are embedding based on proximity rather than structural similarity.

### 6.2 Enron email network

Next we consider an email network encoding email communication between employees in a company. We expect structural equivalences in job titles due to corporate organizational hierarchy.

**Data and setup.** Nodes represent Enron employees and edges correspond to email communication between the employees [21]. An employee has one of seven functions in the company (*e.g.*, CEO, president, manager). These functions provide ground-truth information about roles of the corresponding nodes in the network. We use an embedding learning algorithm to learn an embedding for every Enron employee. We then use these embeddings to compute the average $\ell_2^2$ distance between every two categories of employees.

**Results.** GraphWave captures intricate organizational structure of Enron (Figure 5). For example, CEOs and presidents are structurally distant from all other job titles. This indicates their unique position in the email exchange graph, which can be explained by their local graph connectivity patterns standing out from the others. Traders, on the other hand, appear very far from presidents and are closer to directors. In contrast, *struc2vec* is less successful at revealing intricate relationships between the job titles, yielding an almost uniform distribution of distances between every class. We assess the separation between "top" job titles (CEO and President) and lower levels in the job title hierarchy of all three methods (GraphWave, *struc2vec*, and *RolX*) in Table 3. GraphWave achieves 28% higher homogeneity and 139% higher completeness than *RolX* (and performs even better compared to *struc2vec*).

### 6.3 European airline networks

We next analyze a collection of airline networks encoding direct flights between airports that are operated by different airlines. We expect structural equivalences in airports of different airlines due to the known hub-and-spoke structure, such as Frankfurt (FRA) for Lufthansa and Charles de Gaulle (CDG) for Air France.

**Data and setup.** We consider six airlines operating flights between European airports [2]: 4 commercial (Air France, Easyjet, Lufthansa, and RyanAir) and 2 cargo airlines (TAP Portugal, and European Airline Transport). Each airline is represented with a graph, where nodes represent airports/destinations and edges stand for direct flights between the airports. Altogether, there are 45 airports labeled as: hub airports, regional hubs, commercial hubs, and focus cities,[2] which we use as ground-truth structural roles. For each airline graph, we use a feature learning algorithm to learn an embedding of every airport in the graph. We then stack airport embeddings from different graphs and use them as input to t-SNE (for visualization) and as input to agglomerative clustering (for measuring the homogeneity, completeness, and silhouette score) that returns four airport clusters, indicative of the four ground-truth structural roles.

**Results.** We measure how well airports in each cluster correspond to the ground-truth structural roles. GraphWave outperforms alternative methods on all three clustering metrics (Table 3). It outperforms *RolX* and *struc2vec* by 27% and 24%, respectively (homogeneity), and by 12% and 44% (completeness), with a substantially higher silhouette score.

Figure 6 shows t-SNE visualizations of airport embeddings. We find that *struc2vec* learns embeddings that are dominated by airline networks (note how different shapes are localized in *struc2vec*'s t-SNE plot). In particular, airports from the same airline network, *e.g.*, Ryanair, are trivially embedded close together. This indicates that *struc2vec* cannot generalize embeddings across different airline networks and thus cannot successfully identify which airports are structurally equivalent across airlines (*i.e.*, hub/focus cities for each airline). Further examining the figure, we see that *RolX* learns embeddings whose projections do not exhibit any obvious pattern. In contrast, GraphWave learns embeddings that are indicative of airports' structural equivalences (note how different colors are localized in GraphWave's t-SNE plot). In particular, in GraphWave, airports with the same structural role are embedded close together even if they come from different airline networks, demonstrating GraphWave' ability to learn meaningful structural embeddings for real-world networks.

## 7 CONCLUSION

We have developed a new method for learning structural embeddings in networks. Our approach, GraphWave, uses spectral graph wavelets to generate a structural embedding for each node, which we accomplish by treating the wavelets as a distributions and evaluating the resulting characteristic functions. Considering the wavelets as distributions instead of vectors is a key insight needed to capture structural similarity in graphs.

---

[2]These labels were manually curated from Wikipedia based on M. Garrett: Airport Hubs. *Encyclopedia of Transportation: Social Science and Policy* (2014).

Table 3: Clustering results for Enron and airline networks.

| Dataset | Method | Homogen. | Comple. | Silhouette |
|---|---|---|---|---|
| Enron | RolX | 0.090 | 0.028 | 0.425 |
| | struc2vec | 0.003 | 0.018 | 0.435 |
| | GraphWave | **0.115** | **0.067** | **0.577** |
| Airlines | RolX | 0.244 | 0.326 | -0.054 |
| | struc2vec | 0.250 | 0.254 | -0.035 |
| | GraphWave | **0.310** | **0.365** | **0.050** |

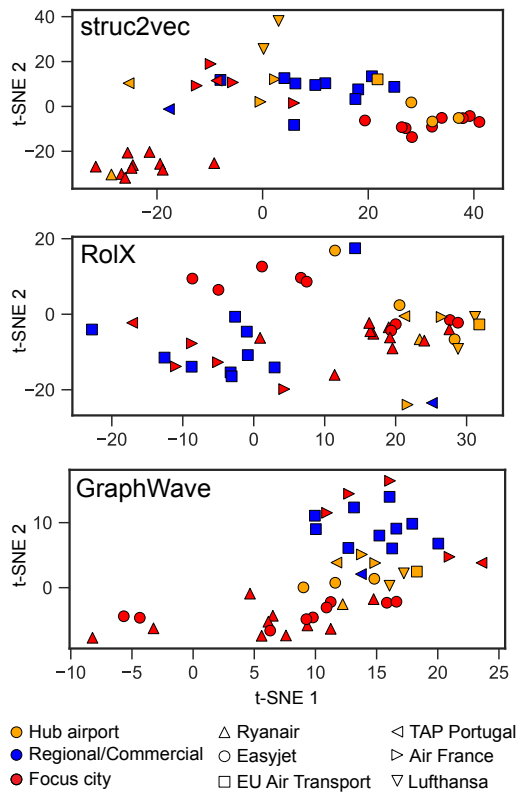

Figure 6: t-SNE projections of airport embeddings.

Our method provides mathematical guarantees on the optimality of learned structural embeddings. Using spectral graph theory, we prove that structurally equivalent (or similar) nodes have near-identical (or similar) embeddings in GraphWave. Various experiments on real and synthetic networks provide empirical evidence for our analytical results and yield large gains in performance over state-of-the-art baselines. For future work, these embeddings could be used for transfer learning, leveraging data from a well-explored region of the graph to infer knowledge about less-explored regions.